\documentclass[chicago,numberedappendix,tighten]{emulateapj}

\usepackage{amsmath,amssymb,natbib,graphicx}
\usepackage{color}
\usepackage{ulem}

\slugcomment{DRAFT \today}

\shorttitle{X-Ray Observations of Supernova 2013ej}
\shortauthors{Chakraborti et al.}

\begin{document}


\title{Probing Final Stages of Stellar Evolution with X-Ray Observations of SN 2013ej}


\author{Sayan Chakraborti\altaffilmark{1}}
\author{Alak Ray\altaffilmark{2}}
\affil{Institute for Theory and Computation, Harvard-Smithsonian Center for Astrophysics, 60 Garden Street, Cambridge, MA 02138, USA}


\author{Randall Smith}
\affil{Harvard-Smithsonian Center for Astrophysics, 60 Garden Street, Cambridge, MA 02138, USA}

\author{Raffaella Margutti}
\affil{Center for Cosmology and Particle Physics, New York University, 4 Washington Place, New York, NY 10003, USA}

\author{David Pooley}
\affil{Department of Physics and Astronomy, Trinity University, San Antonio, TX 78212, USA}

\author{Subhash Bose}
\affil{Aryabhatta Research Institute of Observational Sciences, Manora peak, Nainital, India}

\author{Firoza Sutaria}
\affil{Indian Institute of Astrophysics, Koramangala, Bangalore, India}


\author{Poonam Chandra}
\affil{National Centre for Radio Astrophysics, Pune University Campus, Pune 411 007, India}

\author{Vikram V. Dwarkadas}
\affil{Department of Astronomy and Astrophysics, University of Chicago, 5640 S Ellis Avenue, Chicago, IL 60637, USA}

\author{Stuart Ryder}
\affil{Australian Astronomical Observatory, P.O. Box 915, North Ryde, NSW 1670, Australia}

\author{Keiichi Maeda}
\affil{Department of Astronomy, Kyoto University, Kitashirakawa-Oiwake-cho, Sakyo-ku, Kyoto 606-8502, Japan}


\altaffiltext{1}{Society of Fellows, Harvard University, 78 Mount Auburn Street, Cambridge, MA 02138, USA}
\altaffiltext{2}{Tata Institute of Fundamental Research, 1 Homi Bhabha Road, Colaba, Mumbai 400 005, India}

\email{schakraborti@fas.harvard.edu}



\newpage

\begin{abstract}
Massive stars shape their surroundings with mass loss from winds during
their lifetimes. Fast ejecta from supernovae, from these
massive stars, shocks this circumstellar medium.
Emission generated by this interaction provides a window into the final
stages of stellar evolution, by probing the history of mass loss from the progenitor.
Here we use Chandra and Swift x-ray
observations of the type II-P/L SN 2013ej to probe the history of mass loss from its
progenitor. We model the observed
x-rays as emission from both heated circumstellar matter and supernova
ejecta. The circumstellar
density profile probed by the supernova shock reveals a history of
steady mass loss
during the final 400 years. The inferred mass loss rate of
$3 \times 10^{-6} {\rm \; M_\odot \; yr^{-1}}$ points back to a
14 $M_\odot$ progenitor.
Soon after the explosion we find significant
absorption of reverse shock emission by a cooling shell.
The column depth of this shell observed in absorption provides an independent and consistent
measurement of the circumstellar density seen in emission.
We also determine the efficiency of cosmic ray acceleration from
x-rays produced by Inverse Compton scattering of optical photons
by relativistic electrons. Only about 1 percent of the thermal energy
is used to accelerate electrons.
Our x-ray observations and modeling provides stringent tests
for models of massive stellar evolution and micro-physics of shocks.
\end{abstract}






\section{Introduction}
One of the central problems in astrophysics is the mapping of stellar properties
onto the properties of supernovae that they may or may not produce.
Mass, spin, metallicity, and binarity are some of the parameters which
are thought to determine the final outcome of stellar evolution \citep{2003ApJ...591..288H}.

Type II-P supernovae are produced by red supergiants, between 8 and 17 $M_\odot$ in
mass \citep{2009MNRAS.395.1409S}. X-ray lightcurves of type II-P supernovae
point to an upper limit of 19 $M_\odot$ for their progenitors \citep{2014MNRAS.440.1917D}.
Yet not all stars with such masses necessarily give
rise to type II-P supernovae.
In the final stages of stellar evolution the cores of massive
stars rapidly burn through elements of progressively higher atomic
numbers \citep{1978ApJ...225.1021W}. This may cause rapid variation
in the energy output of the core. However, the outer layers of
these stars need approximately a Kelvin-Helmholtz time
scale $\rm (\sim 10^6 \; yr)$ to adjust to these changes. Therefore, surface properties
like luminosity and mass loss rate, should not change on short timescales in direct response.
However recent observations of luminous outbursts and massive outflows from
Luminous Blue Variable progenitors \citep{2010AJ....139.1451S} months
to years before certain supernovae, like SN 2009ip
\citep{2013MNRAS.430.1801M, 2013Natur.494...65O, 2013ApJ...767....1P, 2014ApJ...780...21M},
call this paradigm into question \citep{2014ARA&A..52..487S}.

The pre-supernova evolution of massive stars
shape their environments by winds and ionizing radiation. The interaction
of the supernova ejecta with this circumstellar matter produces radio and x-ray emission.
Our ongoing program \citep{2012ApJ...761..100C, 2013ApJ...774...30C}
is to observe these x-rays using various sensitive instruments and
model their emission mechanism. In this work, we use Chandra and Swift x-ray
observations of SN 2013ej to probe the history of mass loss from its
progenitor during the last 400 years before explosion.
At early times we find significant
absorption of reverse shock emission by a cooling shell.
We also determine the efficiency of cosmic ray acceleration from
x-rays produced by Inverse Compton scattering of optical photons
by relativistic electrons. Our results demonstrate that
sensitive and timely x-ray observations of young nearby supernovae, coupled
with modeling of the emission and absorption produced by shocked
plasmas, provide stringent tests for models of pre-supernova
massive stellar evolution.

\section{Observations of SN 2013ej}
SN 2013ej exploded in the nearby galaxy M74
\citep{2013CBET.3606....1K,2013CBET.3609....1V,2013CBET.3609....3W,2013CBET.3609....4D}
and was observed in multiple bands.
It was initially classified as a type II-P supernova
\citep{2013ATel.5275....1L,2013ATel.5466....1L,2014JAVSO..42..333R} with
a slow rise \citep{2014MNRAS.438L.101V},
but due to its fast decline \citep{2015ApJ...807...59H}
it was later re-classified as a type II-L supernova \citep{2015MNRAS.448.2608V, 2015ApJ...806..160B}.
\citet{2015ApJ...799..208S} have shown that supernovae of type II-P and II-L form a continuum
of lightcurve properties like plateau duration. SN 2013ej falls somewhere along this
continuum.
In this work we adopt a distance of $d \sim 9.57 \pm 0.70$ Mpc and an explosion
date of 23.8 July 2013 \citep{2015ApJ...806..160B}.
Details of the x-ray observations, carried out by us and used in this work are given below
and in Table \ref{obstab}.

\subsection{Swift XRT Observations}
The Swift XRT observed SN 2013ej in x-ray bands 
starting from 2013 July 30 until 2013 July 31. \citet{2013ATel.5243....1M}
analyzed and reported data collected during the first 15 ks of observations. 
The x-ray counterpart of SN2013ej was found to be separated from nearby sources.
The significance of the x-ray source detection in the Swift observation
was 5.2 sigma. In this work we use x-ray data collected over a longer duration of 73.4 ks
by the XRT.
The X-ray counterpart of SN2013ej, as seen by the XRT, is 45" away from an ULX
source M74 X-1 which is clearly detected and resolved. It is also 15" from an
X-ray source J013649.2+154527 observed by Chandra. These circumstances make
follow-up Chandra observations with superior angular resolution particularly
important.

\subsection{Chandra X-ray Observations}
After the initial detection by Swift,
we triggered our Target of Opportunity observations
with the Chandra X-ray Observatory for 5 epochs. The first Chandra
observation was approximately 10 ks and the subsequent four observations were all
$\sim 40$ ks each. All exposures were carried out using Chandra ACIS-S CCDs without any
grating.
The details of observations of SN 2013ej with Swift and Chandra
x-ray instruments are given in Table \ref{obstab}. 
These data from each epoch of observations were processed separately, but identically.
The spatial and spectral analyses were performed after this initial
processing by following
the prescription\footnote{The method for extraction of spectrum and response files for an unresolved source
is described in \texttt{http://cxc.harvard.edu/ciao/threads/pointlike/}}
from the Chandra Science Center using CIAO 4.7 with CALDB 4.6.9.

The initial data processing steps were identical to that of SN 2004dj \citep{2012ApJ...761..100C}
and SN 2011ja \citep{2013ApJ...774...30C}.
Photons recorded in level 2 events were filtered by energy to select only those above 0.3 keV
and below 10 keV. The selected photons were projected back on to sky coordinates and the
emission from the supernova was easily identified.
The portion of the sky containing the supernova was masked and a light curve was
generated from the remaining counts. Cosmic ray induced flares were identified in this
light curve, using times where the count rate flared $3 \sigma$ above the mean.
A good time interval table was generated by excluding these flares. This was used to
further select photons from the useful exposure times reported in Table \ref{obstab}.
The spectra, response matrices and background count rates were then
generated from these filtered photons. 
To retain the highest available spectral resolution, we did not bin 
these data. All subsequent steps use this processed data.

\begin{figure}
 \includegraphics[angle=0,width=\columnwidth]{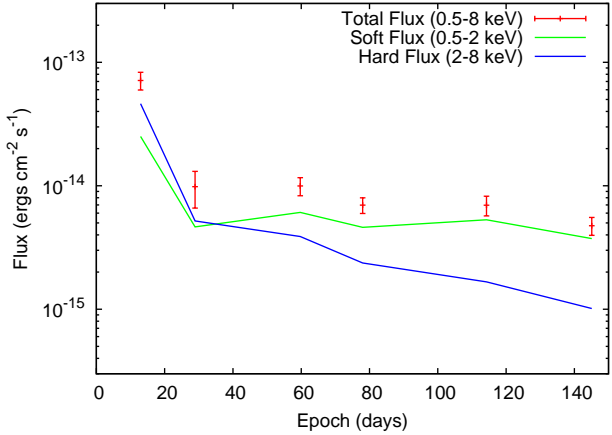}
 \caption{X-ray fluxes observed from SN 2013ej. Total fluxes (red, with $1 \sigma$
 uncertainties) can be split into soft (green) and hard (blue) components.
 Note that the hard component dominates at first. However, it drops off rapidly
 as the Inverse Compton flux dies off and the thermal plasmas become cooler. The soft
 component does not drop off as rapidly, because the reduced emission is somewhat offset
 by reduced absorption at late times.}
 \label{fluxes}
\end{figure}

\section{Modeling the X-Rays}
The expansion of fast supernova ejecta drives a strong forward shock into
circumstellar matter \citep{1974ApJ...188..501C}, and heats it to $\sim 100$ keV.
The expansion also causes rapid adiabatic cooling in the ejecta. However, an
inward propagating reverse shock is generated by the deceleration of the ejecta
by the ambient medium \citep{1974ApJ...188..335M}, reheating it to $\sim 1$ keV or even higher.
We use thermal and non-thermal emission processes, as well as absorption, occurring in
these shocked regions to model x-rays observed (Fig \ref{fluxes}) from SN 2013ej.
Note that the hard x-ray flux, initially the dominant part, rapidly declines
and beyond $\sim40$ days the total flux is dominated by the soft x-ray flux.
The observed spectrum (Fig \ref{ufs}) is represented as the sum of
these emission components, passed through the appropriate absorption components
and folded in with the relevant response matrices. The XSPEC model we used is
$\mathit{tbabs(tbabs(apec)+bremss+powerlaw)}$. Here external absorption is
modeled by the first $\mathit{tbabs}$ and internal by the second one. The $\mathit{apec}$ component
represents thermal emission from reverse shock while the $\mathit{bremss}$ represents
that from forward shock. The Inverse Compton component is represented by $\mathit{powerlaw}$.

\subsection{Thermal Emission}
The reverse shock climbs up against the steep ejecta profile of the supernova
and therefore encounters larger densities than the forward shock. The temperature
of the reverse shock can in many cases be right where Chandra is most
sensitive. Therefore thermal emission from the reverse shock is likely to be
the dominant component at late times beyond a month \citep{2012ApJ...761..100C}.
The thermal emission from the forward shock can become important
if the emission from the reverse shock is absorbed.

The thermal x-rays from the reverse shock are composed of bremsstrahlung
and line emission.
\citet{2006A&A...449..171N} used time-dependent ionization balance and
multilevel calculations to model the line emission from the reverse shock.
\citet{2012ApJ...761..100C} have shown that it is safe to assume
collisional ionization equilibrium while trying to model the line emission
from the reverse shocked material. The strengths of lines from a plasma
in equilibrium can be determined from its temperature and composition.
We use the APEC code \citep{2001ApJ...556L..91S}
to model the thermal emission from the reverse shock.

The thermal emission from the forward shock is modeled simply as bremsstrahlung
radiation with a normalization of $N_{\rm bremss}$.
We expect it to be too tenuous and hot to produce any significant lines in the
Chandra or XRT bands \citep{2012ApJ...761..100C}.

\subsection{Non-Thermal Emission}
The forward shocks, apart from heating the circumstellar material, also
accelerate cosmic rays. The relativistic electrons at the forward shock
lose energy via synchrotron emission, which is detected in the radio
\citep{1982ApJ...259..302C}, and Inverse Compton scattering of optical
photons into the x-rays \citep{2012ApJ...761..100C}.
Here we model the Inverse Compton emission as a power law in XSPEC
with a normalization of $N_{\rm IC}$.
An electron population described by a power law with index $p$ generates
Inverse Compton scattered x-rays with a photon index $(p + 1)/2$.

\begin{figure}
 \includegraphics[angle=0,width=\columnwidth]{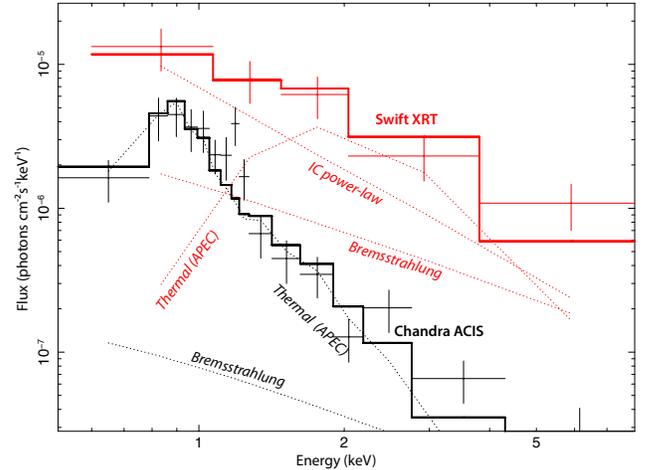}
 \caption{Unfolded x-ray spectra of SN 2013ej from Swift XRT at early times (in red) and Chandra ACIS
 at later times (in black) with $1\sigma$ uncertainties.  The Swift spectrum represents the earliest epoch
 and has comparable contribution from thermal reverse shock emission, thermal forward shock
 emission and Inverse Compton scattering. The later time Chandra spectra from 5 epochs listed
 in Table \ref{obstab}, are stacked together only for display but analyzed separately.
 Note that the late time Chandra spectra are softer than the early time Swift spectrum.
 The Chandra spectra are dominated by thermal emission from the reverse shock.}
 \label{ufs}
\end{figure}

\subsection{Absorption components}
We consider two absorption components, both modeled using the
Tuebingen-Boulder ISM absorption model \citep{2000ApJ...542..914W}.
We consider the external absorption to be a constant in time as it
is likely produced by material far away from the supernova.
Radiative cooling of the reverse shocked material
leads to the formation of a dense cool shell \citep{2003LNP...598..171C}
which can obscure the emission from the reverse shock. We model this
as a time-varying internal absorption component.

\begin{deluxetable*}{l r c l r c}
\tablecolumns{6} 
\tablewidth{0pc} 
\tablecaption{Swift and Chandra observations of SN 2013ej} 
\tablehead{ 
Dates  & \multicolumn{1}{c}{$\rm Age$\tablenotemark{a}} & $L_{\rm bol}$ & Telescope & Exposure & X-ray Flux (0.5-8 keV)\\
(2013) & (days)         & (ergs s$^{-1}$) &           & (ks) & (ergs cm$^{-2}$ s$^{-1}$)}
\startdata 
Jul 30 - Aug 9 & 13.0 & $(3.89 \pm 0.58) \times10^{42}$ & Swift  & 73.4 & $(7.1 \pm 1.2) \times10^{-14}$ \\
Aug 21         & 28.9 & $(2.19 \pm 0.27) \times10^{42}$ & Chandra&  9.8 & $(9.8 \pm 3.2) \times10^{-15}$ \\
Sep 21         & 59.7 & $(1.29 \pm 0.03) \times10^{42}$ & Chandra& 39.6 & $(1.0 \pm 0.2) \times10^{-14}$ \\
Oct 7 - 11     & 78.0 & $(1.00 \pm 0.02) \times10^{42}$ & Chandra& 38.4 & $(7.0 \pm 1.0) \times10^{-15}$ \\
Nov 14         & 114.3 & $(8.13 \pm 0.38) \times10^{40}$ & Chandra& 37.6 & $(7.0 \pm 1.3) \times10^{-15}$ \\
Dec 15         & 145.1 & $(5.13 \pm 0.24) \times10^{40}$ & Chandra& 40.4 & $(4.8 \pm 0.8) \times10^{-15}$ \\
\label{obstab}
\enddata

\tablecomments{The Chandra observations can be retrieved from the Chandra Data Archive
using their Obs Ids of 14801, 16000, 16001
(with fragments in 16484 and 16485), 16002, and 16003.}
\tablenotetext{a}{Age at the middle of an observation with
an assumed explosion date 23.8 July 2013 (UT) (JD 2456497.3 $\pm$ 0.3)
following \citet{2015ApJ...806..160B}}
\end{deluxetable*}

\section{X-ray Spectral Fitting}
All x-ray data are loaded into XSPEC and fitted in the manner described in
\citet{2012ApJ...761..100C}. Since these data are unbinned, individual
spectral channels can have a low number of photons, disallowing the use
of a $\chi^2$ statistic. We therefore adopt the $W$ statistic generalization of
the \citet{1979ApJ...228..939C} statistic.
We need to fit 6 epochs with 10 parameters each. Since there
is not enough information in the observed spectra to simultaneously
determine all 60 parameters, it is necessary to hold some of them
constant or constrained. We describe these restricted parameters below.
All fitted parameters are reported in Table \ref{fitparms}.

\subsection{Constant parameters}
\label{constant}
\citet{2015ApJ...806..160B} find no excess reddenning in the optical
emission from SN 2013ej beyond what is expected from the Galactic
absorption. We therefore hold the external absorption column constant, at the Galactic value
of $n_{\rm ext} = 4.8 \times 10^{20}$ atoms cm$^{-2}$
determined from the Leiden Argentine Bonn (LAB) Survey of Galactic HI
\citep{2005A&A...440..775K}.

A visual inspection of the spectra reveals a bump at $\sim 1$ keV, which is
likely produced by a blend of lines, but not enough resolved features
to determine the metallicity of the plasma.
We therefore set the relative metal abundances in APEC following \citet{2009ARA&A..47..481A}.
The overall metallicity is set to $Z=0.295 Z_\odot$, which is equal to that of the nearby HII
region number 197 of \citet{2012A&A...545A..43C}.
In the absence of prominent sharply resolved lines, the redshift cannot be determined
from the spectra. We therefore fix it to the host galaxy redshift of $z=2.192\times 10^{-3}$ from NED.

The early spectrum at the first epoch is hard, with a possible contribution from Inverse Compton
scattering. But there is unlikely to be enough information to be able to determine
the slope of this component. We therefore fix the photon index to $\alpha_{\rm IC}=2$, which is
expected on theoretical grounds for an electron index of $p=3$ and has been
observed in SN 2004dj \citep{2012ApJ...761..100C}.

\subsection{Constrained parameters}
\label{constrainted}
Here we constrain various parameters which determine how the shape of the
spectra change in time. To derive these relations, we assume a steady
mass loss rate from the progenitor. If the mass loss is significantly
variable, the data will rule out the model.
We allow the absorption column depth of the cool shell to be
determined by the best-fit to these data. However, the value of relative
depth of the column at 6 epochs are tied to each other using
the relation
\begin{equation}
n_{\rm cool} \propto t^{-1}
\end{equation}
from \citet{2003LNP...598..171C}.
This removes 5 free parameters.

At each epoch the temperature of the forward shock can be related to that
of the reverse shock. Using the self similar solution for a supernova
ejecta interacting with a steady wind \citep{1982ApJ...258..790C}, we
find,
\begin{equation}
T_{\rm cs} = (n-3)^2 T_{\rm rev},
\end{equation}
where $n$ is the power law index of the ejecta profile.  Following
\citet{1999ApJ...510..379M} we use $n=12$, as is appropriate for
a red supergiant progenitor. Fixing the forward shock temperature
to be $81$ times the reverse shock temperature at each of the epochs
removes 6 free parameters. The temperature of the reverse shocked plasma
also goes down slowly in time, as
\begin{equation}
T_{\rm rev} \propto t^{-\frac{2}{n-2}}.
\end{equation}
The temperature of the $\mathit{apec}$ component at one epoch is therefore allowed to vary
but its values at all other epochs are linked to each other using this
proportionality. This removes 5 more free parameters.

The emission measures of the plasma at the forward and reverse shocks
can be similarly related. Self similar solutions \citep{2003LNP...598..171C}
provide the physical relation between the emission measures as
\begin{equation}
 \int n_e n_H dV_{\rm rev} = \frac{(n-3)(n-4)^2}{4 (n-2)} \int n_e n_H dV_{\rm cs}.
\end{equation}
Two more factors
arise because we are forced to use two different models for the emissions,
namely APEC and bremsstrahlung. In XSPEC, the APEC model \citep{2001ApJ...556L..91S}
represents the emission measure as $\int n_e n_H dV$, whereas the $\mathit{bremss}$
model \citep{1975ApJ...199..299K} uses $\int n_e n_I dV$.
To resolve this, we approximate $n_I=n_H+n_{He}$ with the Helium abundance from
\citet{2009ARA&A..47..481A}. Furthermore, there is an arbitrary numerical difference in the
normalizations of the models. Accounting for these three issues, we set the
$\mathit{bremss}$ norm $N_{\rm bremss}$
to be $0.0228$ times the $\mathit{apec}$ norm $N_{\rm APEC}$ at each epoch. This eliminates another 6 free parameters.

Only the earliest epoch is likely to have significant contribution from
Inverse Compton scattering of optical photons by relativistic electrons.
\citet{2006ApJ...651..381C} have shown that
the Inverse Compton flux, is expected to fall off as
\begin{equation}
 N_{\rm IC} \equiv E \frac{dL_{\rm IC}}{dE} \propto L_{\rm bol} t^{-1},
\end{equation}
where $L_{\rm bol}$ is the bolometric luminosity of the supernova
which provides the seed photons to be up-scattered. We relate the norm
of the $\mathit{powerlaw}$ component, representing the Inverse Compton emission,
at all later epochs to that of the first epoch using this relation.
To estimate the bolometric luminosity before 30 days, we use the B
and V band luminosities from \citet{2014JAVSO..42..333R} and the bolometric
correction prescribed by \citet{2009ApJ...701..200B}. Beyond 30 days,
we use the bolometric luminosity reported in \citet{2015ApJ...806..160B}
by integrating the emission from the infrared to ultraviolet.
All the bolometric luminosities used are reported in Table \ref{obstab}.

\subsection{Goodness and Uncertainties}
Having obtained the best fit, we tested the goodness of the fit
by generating a set of 1000 simulated spectra, at each epoch,
with a parameter distribution that is derived from the covariance matrix of parameters
at the best fit.
We note that goodness testing is a misnomer for this process as it can never determine
whether a particular fit is good, only if it is significantly bad or not.
Only 60 percent of these sets of fake data have a fit statistic better than the
fiducial fit.
If the observations were indeed generated by the model the most likely
percentage, of fake data that have a fit statistic better than the
fiducial fit, is 50. However, the likelihood of the percentage lying outside
the range of 40 to 60 percent, is $0.8$.
Since the outcome of the goodness test is quite likely, our data do not
rule out the model. We therefore consider the model to be acceptable.

\begin{figure}
 \includegraphics[angle=0,width=\columnwidth]{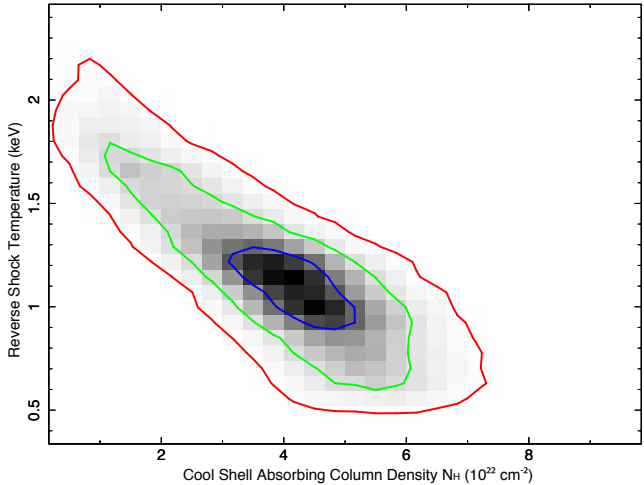}
 \caption{Correlation between cool shell absorption column depth and reverse shock
 plasma temperature at the first epoch. 1 (blue), 2 (green) and 3 (red) $\sigma$
 uncertainty contours are obtained by
 marginalizing the results of our MCMC run. A larger absorbing column can hide lower energy
 emission from a cooler reverse shocked plasma, giving rise to the negative correlation.
 The closed $3 \sigma$ contour demonstrates that even with the uncertainty in the reverse shock temperature
 we need a non-zero column depth in the cooling shell absorption component at the $3 \sigma$.}
 \label{revtemp_shellabs}
\end{figure}

In order to better understand the uncertainties in the determined parameters
we ran a Markov Chain Monte Carlo simulation. 200 walkers were initiated
using the fit covariance matrix as the proposal distribution. They were allowed
to walk for 400,000 steps, after rejecting the first 40,000 steps. They were
evolved following
the Goodman-Weare algorithm
\citep{2012ApJ...745..198H} implemented in XSPEC \citep{2013HEAD...1311704A}.
The uncertainties for each parameter were determined
by marginalizing over all other parameters. Two pairs of parameters were
found to have noteworthy correlations and are discussed below.

\begin{figure}
 \includegraphics[angle=0,width=\columnwidth]{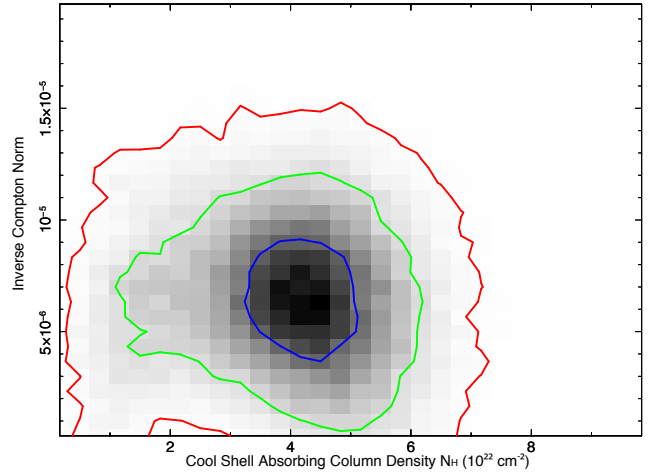}
 \caption{Correlation between cool shell absorption column depth and Inverse Compton
 emission at the first epoch. 1 (blue), 2 (green) and 3 (red) $\sigma$ uncertainty contours
 are obtained by
 marginalizing the results of our MCMC run. A larger absorbing column can hide reverse shock
 emission allowing harder forward shock emission to dominate the spectrum. This makes
 the thermal contribution to the spectrum harder and therefore requires less Inverse Compton
 emission to explain the high energy photons, giving rise to the negative correlation.
 The closed $2 \sigma$ contour demonstrates that after marginalizing over the uncertainties
 in the thermal components, we need a non-zero contribution from the non-thermal Inverse
 Compton component at the $2 \sigma$ level.}
 \label{ic_shellabs}
\end{figure}

The uncertainty in the column depth of the cool shell influences the uncertainties in
the reverse shock temperature (see Figure \ref{revtemp_shellabs}) and the Inverse
Compton flux density (see Figure \ref{ic_shellabs}). A heavier absorbing column
can hide the lower energy emission from a colder plasma, leading to the negative
correlation with the reverse shock temperature. A heavier absorbing column, having
hidden much of the reverse shock emission, also allows for a larger hot bremsstrahlung
contribution from the forward shock. This explains away more of the harder photons,
thus requiring a lesser contribution from the Inverse Compton component. This causes
the column depth to be also negatively correlated with the Inverse Compton flux.

\section{Results}
We interpret the plasma parameters determined from the model fits to
our observations in terms of a physical description of the supernova
and its progenitor.

\subsection{Shock velocity}
\citet{1982ApJ...259..302C} related the temperature of the forward shocked
material with the shock velocity. \citet{2006A&A...449..171N} used this to derive
the temperature of the reverse-shocked material in terms of the forward shock velocity.
The reverse shocked plasma is expected to be dense enough \citep{2012ApJ...761..100C}
to reach ionization equilibrium. Under such conditions,
\citet{2012ApJ...761..100C} have inverted this relation to express the forward shock velocity,
which is a property of the supernova explosion, to the reverse shock temperature
which is an observable, as
\begin{equation}
 V_{\rm cs} = 10^4 \; \sqrt{\frac{kT_{\rm rev}}{1.19 \; {\rm keV}}} \; \; {\rm km \; s^{-1}}.
\end{equation}
Since the best-fit temperature is $1.1\pm0.2$ keV (see Fig \ref{revtemp_shellabs}),
the implied velocity at 12.96 days is $V_{\rm cs}=(9.7 \pm 1.1)\times 10^3$ km s$^{-1}$.
This is faster than the shock velocity observed in SN 2004dj \citep{2012ApJ...761..100C}.
Also, note that the forward shock is expected to be faster than the photosphere. As
expected, $V_{\rm cs}$ here is faster than velocities seen in optical line-widths
\citep{2015ApJ...806..160B}.

\begin{figure}
 \includegraphics[angle=0,width=\columnwidth]{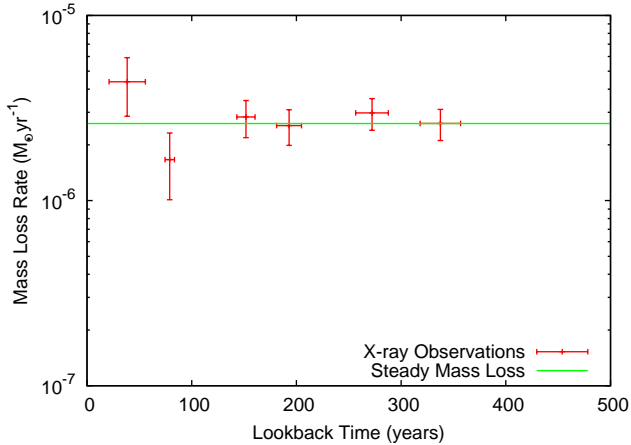}
 \caption{Pre supernova mass loss rate from the progenitor as a function of time before
 explosion with $1\sigma$ uncertainties. The mass loss rates are derived using thermal emission from shocked plasma
 measured in the x-rays. Note that our measurements are consistent with a steady mass loss
 rate of $\dot{M}=(2.6 \pm 0.2) \times 10^{-6} {\rm \; M_\odot \; yr^{-1}}$ for the
 last 400 years of pre-supernova stellar evolution.}
 \label{mdot_tminus}
\end{figure}

\subsection{Mass loss history}
Mass loss from the progenitor sets up the circumstellar density within which the supernova
interacts. The circumstellar density determines the emission measure of the forward
shocked material. This can be related to the emission measure of the reverse shocked material
using self similar solutions \citep{2012ApJ...761..100C}. Considering only the contribution
of Hydrogen and Helium to the mass and number density of the outermost shells of the supernova,
$\rho= 1.17 {\rm\; amu} \times n_e = 1.34 {\rm\; amu} \times n_H$. Therefore we modify
the emission measure derived by \citet{2012ApJ...761..100C} for the reverse-shocked material as
\begin{align}\label{em}
 \int n_e n_H dV =\frac{ (144 / \pi) \left( \dot M / v_{\rm w}\right)^2 }{(1.17 {\rm\; amu})(1.34 {\rm\; amu})R_{\rm cs} }
\end{align}
Only half of this emission measure contributes to the observed flux, as the other half
is absorbed by the opaque unshocked ejecta. The norm of $\mathit{apec}$ in XSPEC is defined as
\begin{equation}
 N_{\rm APEC} = \frac{10^{-14}}{4 \pi \left( D_{\rm A} (1+z)\right)^2} \int n_e n_H dV,
\end{equation}
where $D_{\rm A}$ is the angular diameter distance to the source.
Therefore, the mass loss rate can now be determined
from the emission measure as
\begin{align}
 \dot{M}=&7.5\times10^{-7} \left(\frac{v_{\rm w}}{10 {\rm \; km \; s^{-1}}}\right)
   \left(\frac{D_{\rm A} (1+z)}{10 {\rm \; Mpc}}\right) \nonumber \\
 &\times \left(\frac{N_{\rm APEC}}{10^{-5}}\right)^{1/2}
 \left(\frac{R_{\rm cs}}{10^{15} {\rm \; cm}}\right)^{1/2} {\rm M_\odot \; yr^{-1}}.
\end{align}
We determine the norm of the APEC emission measure directly from our fit. We calculate the
radius from the velocity determined in the last section and the time of observation.
We assume a wind velocity $v_{\rm w} = 10$ km s$^{-1}$ as is appropriate for red supergiant
progenitors.

The emission measures determined at various times after the explosion, point back
to mass loss rates at different lookback times before the explosion. These are plotted
in Figure \ref{mdot_tminus} as a function of the lookback time $t_{\rm look}$.
Note that our observations are consistent with a $\propto r^{-2}$ density
profile as expected from a steady
mass loss rate of $\dot{M}=(2.6 \pm 0.2) \times 10^{-6} \times (v_{\rm w} / 10 {\rm \; km \; s^{-1}}) {\rm \; M_\odot \; yr^{-1}}$
over the last $400 \times (v_{\rm w} / 10 {\rm \; km \; s^{-1}})^{-1}$ years of pre-supernova stellar evolution.

\begin{figure}
 \includegraphics[angle=0,width=\columnwidth]{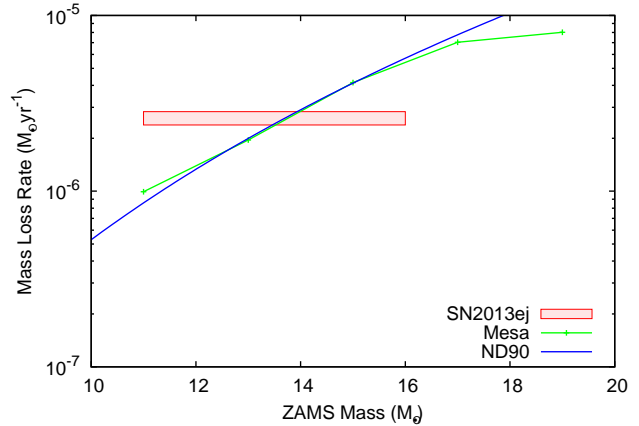}
 \caption{Zero Age Main Sequence Mass (ZAMS) and wind mass loss rate for MESA runs
 (green), and theoretical  line (blue) from \citet{1990A&A...231..134N}
 plotted  for comparison. Shaded box represents $1\sigma$
 confidence intervals for mass loss rate observed in SN 2013ej
 ($\dot{M}=(2.6 \pm 0.2) \times 10^{-6} {\rm \; M_\odot \; yr^{-1}}$, see Fig \ref{mdot_tminus}
 of this work)
 and the estimated progenitor mass (${\rm11-16\; M_\odot}$, from literature). Note that the observed
 mass loss is consistent with the theoretical expectations of mass loss rates from
 red supergiant stars \citep{1990A&A...231..134N}. Our mass loss rate
 measurement points back to a more precise estimate of the progenitor mass,
 $M_{\rm ZAMS} = 13.7 \pm 0.3 {\rm \; M_\odot}$.}
 \label{mmdot}
\end{figure}

We compare this observed mass loss rate of the progenitor
of SN 2013ej, with what is expected from theory.
MESA \citep{2011ApJS..192....3P} was used to simulate
stars with masses between 11 to 19 $M_{\odot}$, for half solar metallicity.
\citet{2011ApJS..192....3P} Sec 6.6 describe the mass loss
prescription used in our simulations as the {\it Dutch} Scheme.
We expect the supernova ejecta to encounter the mass
lost during the RGB phase of the wind which follows \citet{1988A&AS...72..259D}.
In Figure \ref{mmdot} we compare the observed mass loss rate with those
obtained from MESA and from \citet{1990A&A...231..134N} for a progenitor
size of $10^3R_\odot$. Note however that various modifications
have been suggested to this prescription \citep{2011A&A...526A.156M}.

We consider a progenitor mass ranging from 11 to 16 ${\rm M_\odot}$
combining pre-supernova progenitor identification
\citep{2014MNRAS.439L..56F} and modeling of the supernova
lightcurve \citep{2015ApJ...806..160B,2015arXiv150901721D}. Within this mass range,
the mass loss rate obtained from x-ray observations in this work are in
agreement with the predictions from both MESA and \citet{1990A&A...231..134N}.

\begin{deluxetable*}{rccccccccc} 
\tablecolumns{10} 
\tablewidth{0pc} 
\tablecaption{Spectral Fits to X-ray Observations of SN 2013ej} 
\tablehead{ 
\colhead{}    &  \multicolumn{6}{c}{Model Parameters} &   \colhead{}   & 
\multicolumn{2}{c}{Derived Quantities} \\ 
\cline{2-7} \cline{9-10} \\ 
\colhead{$t_{\rm obs}$} &
\colhead{$N_{\rm IC}$} & \colhead{$T_{\rm cs}$} & \colhead{$N_{\rm bremss}$} &
\colhead{$n_{\rm cool}$}    & \colhead{$T_{\rm rev}$} & \colhead{$N_{\rm APEC}$} & 
\colhead{}    & \colhead{$t_{\rm look}$}    & \colhead{$\dot{M}$} \\
\colhead{days}  & \colhead{${\rm ph \; keV^{-1}cm^{-2}s^{-1}}$} & \colhead{keV}
& \colhead{} & \colhead{$10^{22}{\rm \; cm^{-2}}$} & \colhead{keV} & \colhead{$10^{-5}$} & 
& \colhead{yr} & \colhead{$10^{-6} {\rm \; M_\odot \; yr^{-1}}$}
}
\startdata 
$13.0\pm5.8$ & $(6.5\pm2.7)\times 10^{-6}$ & \nodata & \nodata
& $4 \pm 1 $ & $1.13 \pm 0.25$ & $30.8 \pm 15.7$ & 
& $\phn 48 \pm 17$ & $4.38 \pm 1.53$ \\
$28.9\pm0.3$ & \nodata & \nodata & \nodata
& \nodata  & \nodata & $2.16 \pm 1.66 $ & 
& $\phn 79 \pm 5\phn$ & $1.67 \pm 0.65$ \\
$59.7\pm0.4$ & \nodata & \nodata & \nodata
& \nodata  & \nodata & $3.25 \pm 1.37 $ & 
& $152 \pm 9\phn$ & $2.83 \pm 0.64$ \\
$78.0\pm1.8$ & \nodata & \nodata & \nodata
&  \nodata  &  \nodata  & $2.06 \pm 0.82$ & 
& $193 \pm 12$ & $2.54 \pm 0.55$ \\
$114.3\pm0.4$ & \nodata & \nodata & \nodata
& \nodata  & \nodata & $2.00 \pm 0.70 $ & 
& $272 \pm 16$ & $2.98 \pm 0.58$ \\
$145.1\pm0.4$ & \nodata & \nodata & \nodata
& \nodata  & \nodata & $0.82 \pm 0.43 $ & 
& $338 \pm 19$ & $2.61 \pm 0.50$ \\
\cutinhead{Scalings for \ldots entries}
  & $\propto L_{\rm bol} t^{-1}$ & $=81T_{\rm rev}$ & $= 0.0228 N_{\rm APEC}$
&  $\propto t^{-1}$  &  $\propto t^{-1/5}$  &  & 
&  & 
\label{fitparms}
\enddata 
\tablecomments{Free model parameters are determined from fits. Entries marked
with \ldots are not frozen. They are allowed to vary, but only
in proportion to other parameters as described in Section \ref{constrainted}.
Apart from the 6 columns of model parameters listed above, each epoch also has
4 other parameters which are held constant and are identical at each epoch.
These parameters are the external absorbing column density
$n_{\rm ext}= 4.8 \times 10^{20} {\rm \; atoms \; cm^{-2}}$, powerlaw slope for the Inverse Compton
component $\alpha_{\rm IC}=2$, redshift $z=2.192\times 10^{-3}$, and metallicity $Z=0.295 Z_\odot$.
The motivations for fixing the parameters to these particular values are described in
Section \ref{constant}. The norms for the $\mathit{bremss}$ and $\mathit{apec}$ components are reported in the
units used inside XSPEC, so that readers can reproduce the model easily.
The lookback time and mass loss rates are derived from the model parameters, for a
progenitor wind velocity of
$v_{\rm w} = 10$ km s$^{-1}$.}
\end{deluxetable*}

\subsection{Cooling shell absorption}
\citet{2003LNP...598..171C} proposed that a shell of material formed by the
radiative cooling of shocked material may form between the reverse and forward
shocked materials. Though more material is cooled with time, it gets diluted
with the expansion of the ejecta. Also, as the density of the reverse shocked
material falls, it does not cool as effectively as before.
Therefore, this shell poses larger
absorbing column densities at early times. Since the emission from the reverse
shocked material is softer, hiding some of it makes the total spectrum harder.
We determine the column density of this cold material at 12.96 days to be
$n_{\rm cool} = (4 \pm 1) \times 10^{22} {\rm \; atoms \; cm^{-2}}$. This is enough
to block most of the reverse shock emission at early times. This level of
variable absorption is at tension ($\sim 2 \sigma$ level) with the
expected value \citep{2003LNP...598..171C}. This could be the result
of excess absorption from partially ionized wind in the circumstellar material.

The amount of material in the cool shell depends upon the density of the ejecta
which the reverse shock runs into. In a self similar explosion this depends on the
circumstellar density
and hence the mass loss rate from the progenitor \citep{2003LNP...598..171C}.
We find that the observed column density of cool material may be explained
by a mass loss rate of $\dot{M}=(6 \pm 3) \times 10^{-6} {\rm \; M_\odot \; yr^{-1}}$.
This is less precise than, but consistent with, the mass loss rate derived
from the emission measure. This provides a consistency check for the scenario
in which the excess absorption at early times indeed arises from the cooling shell.

\subsection{Particle acceleration}
Electrons are accelerated in the strong forward shock produced by the supernova.
The optical photons produced by the supernova are Inverse Compton scattered
into the x-ray band by these relativistic electrons. Our measurement of the
Inverse Compton flux density provides a direct probe of the particle acceleration
efficiency.

Following \citet{2006ApJ...651..381C}, we can express the Inverse Compton flux,
for an electron index of $p=3$, as
\begin{align}
 E\frac{dL_{\rm IC}}{dE}&\approx8.8\times10^{37} \gamma_{\rm min} \epsilon_{\rm e}
 \left( \frac{\dot{M}/(4 \pi v_{\rm w})}{5 \times 10^{11} {\rm \; g \; cm^{-1}}} \right) \nonumber \\
 &\times \left( \frac{V_{\rm cs}}{10^4 {\rm \; km \; s^{-1}}} \right) 
 \left(\frac{L_{\rm bol}}{10^{42} {\rm \; ergs \; s^{-1}}}\right) \nonumber \\
 &\times \left(\frac{t}{10 {\rm \; days}} \right)^{-1} {\rm \; ergs \; s^{-1}},
\end{align}
where $\gamma_{\rm min}$ is the minimum Lorentz factor of the relativistic electrons
and $\epsilon_{\rm e}$ is the fraction of thermal energy given to relativistic electrons.
Our measurement of the Inverse Compton flux density implies an electron acceleration
efficiency of $\gamma_{\rm min}\epsilon_{\rm e}=0.02\pm0.01$. This shows that
for a $\gamma_{\rm min}=2$, around $1\%$ of the thermal energy is used to
accelerate relativistic electrons.

\section{Discussion}
Explosions of massive stars with extended hydrogen envelopes produce
Type II supernovae. The cores of these stars undergo rapid evolution
during the final millennium before collapse, as they burn elements
with progressively higher atomic numbers. The outer layers of
these stars, supported against gravity by the energy generation in
the core, can only slowly adjust to these changes over a much longer Kelvin-Helmholtz time
scale. Therefore, conditions at the surface of the star, including
luminosity and mass loss rate, are not expected to reflect the rapid
evolution taking place in the core during the last stages of stellar
evolution. This paradigm has been called into question by recent
observations of luminous outbursts and massive outflows observed months
to years before certain supernovae.

Our x-ray observations of SN 2013ej indicate a mass loss rate from the progenitor
which remained steady in the last 400 years before explosion. Within the best
constraints the mass loss rate is consistent
with stellar evolution models and theoretical mass loss prescriptions.
If theoretical mass loss rate predictions are to be trusted, our
precise measurement of the mass loss rate can be used to derive a mass
of $M_{\rm ZAMS} = 13.7 \pm 0.3 {\rm \; M_\odot}$. The statistical uncertainty
in such a measurement rivals the most precise progenitor mass measurements.
However, we need to address gaps in our understanding of mass loss from massive
stars \citep{2014ARA&A..52..487S} before we can quantify systematic errors
and rely on the accuracy of such a measurement.
The mass loss rate inferred here is larger than that observed by us in the Type II-P SN 2004dj
\citep{2012ApJ...761..100C}. The mass loss rate from the progenitor of the Type II-P SN 2011ja
showed rapid variations in the final stages before explosion
\citep{2013ApJ...774...30C,2015arXiv150906379A}.
No such variation is inferred for SN 2013ej and its steady mass loss rate
is comparable to the higher end of mass loss rates inferred for SN 2011ja.
Through our program of x-ray observations of nearby supernovae, we hope to shed
light on details of mass loss from massive stars both as a function of
progenitor mass and lookback time before explosion.

SN 2013ej was caught much sooner after explosion than SN 2004dj or SN 2011ja
thanks to timely Swift and Chandra observations. This allowed us to discover
two interesting effects. \citet{2003LNP...598..171C} postulated the presence
of a cool shell which may obscure the reverse shock emission at early times.
We not only see this effect but measure the column depth of this shell and
confirm that it is consistent with the circumstellar density seen in emission.
If we can measure this effect more precisely in the future, the combination
of the same mass loss rate measured using absorption and emission may allow
an independent determination of the distance to nearby supernovae.
\citet{2006ApJ...651..381C} had suggested that Inverse Compton scattering
by relativistic electrons may be the dominant source of x-rays in some supernovae.
At early times, when the light of the SN 2013ej provides a bright source of
seed photons, emission from this non thermal process is found to be comparable to 
those from thermal processes (see early XRT spectrum in Fig \ref{ufs}). We use this to measure
the efficiency of relativistic electron acceleration. Our measurement
provides a check for recent predictions of particle acceleration
efficiencies in strong but non-relativistic shocks
\citep{2007ApJ...661..879E,2015ApJ...809...55B,2015PhRvL.114h5003P}.

We have also considered the detectability of core collapse supernovae in external
galaxies in the harder X-ray bands. With the capability of NuSTAR
\citep{2013ApJ...770..103H}, SN 2013ej would
have been detected in  6-10 keV band at a $3\sigma$ level with an exposure of 1 Ms,
provided the SN was targeted immediately after discovery and classification.
Thus only very young and very nearby supernovae, e.g. within $2-3\; \rm Mpc$ can
be realistically targeted for detections in the high energy bands in the near future.

\acknowledgments
We acknowledge the use of public data from the Swift data archive.
This research has made use of data obtained using the Chandra X-ray Observatory
through an advance Target of Opportunity program and
software provided by the Chandra X-ray Center (CXC) in the application packages CIAO
and ChIPS.
Support for this work was provided by the National Aeronautics and Space Administration
through Chandra Award Number G04-15076X issued by the Chandra X-ray Observatory Center,
which is operated by the Smithsonian Astrophysical Observatory for and on behalf of the
National Aeronautics Space Administration under contract NAS8-03060.
We thank Naveen Yadav for MESA runs and the anonymous referee for useful suggestions. 
A.R. thanks the Fulbright Foundation for a Fulbright-Nehru Fellowship
at Institute for Theory and Computation (ITC), Harvard University,
and the Director and staff of ITC for their hospitality during his
sabbatical leave from Tata Institute of Fundamental Research. At Tata 
Institute this research is part of 12th Five Year Plan Project 12P-0261.










\bibliographystyle{hapj}
\bibliography{master}

\end{document}